\documentclass[12pt,a4paper]{article}
\usepackage[usenames, dvipsnames]{xcolor}
\usepackage[normalem]{ulem}
\usepackage{jcapmod}

\usepackage{tocloft,ulem}
\usepackage[]{todonotes}
\usepackage{verbatim}
\usepackage{mathrsfs}
\usepackage{cleveref}
\usepackage[shortlabels]{enumitem}
\usepackage{pgf}
\usepackage{tikz-cd}
\usetikzlibrary{shapes,arrows}
\usetikzlibrary{calc}
\usepackage{pgfplots}
\pgfplotsset{compat=1.12}
\usepackage{tikz-3dplot}
\usepackage{slashed}

\usepackage{amsthm}

\usepackage{tikz}
\usepackage{setspace,caption}
\usepackage{lipsum}
\usetikzlibrary{matrix,arrows,calc}

\usepackage{amsmath}
\usepackage{amssymb}
\usepackage{amsfonts}
\usepackage{mathtools}

\newcommand{\I}{\mathrm{i}}

\newcommand{\dif}{\,\mathrm{d}}

\usepackage{comment}

\definecolor{darkgreen}{rgb}{0.23, 0.56, 0.22}

\newcommand{\ho}[0]{\,{\rm hol}}

\title{LVS Scalings}

\begin{document}
	\newcommand{\main}{.}
\begin{titlepage}

\setcounter{page}{1} \baselineskip=15.5pt \thispagestyle{empty}
\setcounter{tocdepth}{2}
\bigskip\

\vspace{.1cm}
\begin{center}
{\Large \bfseries An attempt to derive the LVS\\[.5cm] from a 10d EFT perspective}
\end{center}

\vspace{0.55cm}

\begin{center}
\scalebox{0.95}[0.95]{{\fontsize{14}{30}\selectfont 
Arthur Hebecker,$^{a}$ Andreas Schachner,$^{b}$ and Gaetano Maria Sifo$^{a}$\vspace{0.25cm}}}

\end{center}

\begin{center}

\vspace{0.15 cm}
{\fontsize{11}{30}
\textsl{$^{a}$Institute for Theoretical Physics, Heidelberg University, Philosophenweg 19, 69120 Heidelberg, Germany}\\[.1cm]
\textsl{$^{b}$Department of Physics, Cornell University, Ithaca, NY 14853 USA}\\
}
\vspace{0.25cm}

\vspace{1.1cm}
\normalsize{\bf Abstract} \\[8mm]
\end{center}

\begin{center}
    \begin{minipage}[h]{15.0cm}
   Non-perturbative scalar potentials are a central ingredient for stabilising moduli in flux compactifications. A key target in this context is the direct 10d derivation of the LVS AdS potential, i.e.,~the scalar potential of the Large Volume Scenario before the uplift. The familiar result, relying on 4d supergravity machinery and large volume expansion, consists of three terms of parametric form $\exp(-2a\tau_s)/{\cal V}$, $W_0\exp(-a\tau_s)/{\cal V}^2$, and $W_0^2/{\cal V}^3$. In plain 10d EFT language, the first term is an instanton-anti-instanton effect while the second term comes from a single instanton coupled to 3-form flux.
    When attempting to derive the suppression by the volume ${\cal V}$ directly from the 10d EFT, one encounters a puzzle:
    One expects a flux-dilution factor $1/\sqrt{\cal V}$ for the second term and a factor $1/{\cal V}^2$ from Weyl rescaling to the 4d Einstein frame for both the first and second term. 
    Thus, the total volume suppression for these two terms is stronger than in the aforementioned 4d EFT expectations.
     We discuss the underlying logic in detail, also in the gaugino condensation case, but fail to find a satisfactory explanation for the mismatch. While log-corrections to the K\"ahler potential may be responsible, these would have to be strong enough to affect the outcome of the LVS scheme at the ${\cal O}(1)$ level. What is worse, the origin of the required correction remains unclear.
    
    \end{minipage}

\end{center}

\end{titlepage}
\tableofcontents

\hrulefill

\section{Introduction}

Realising meta-stable de Sitter vacua in string theory is a difficult undertaking, with the most explicit and best studied proposals being the KKLT \cite{Kachru:2003aw} and LVS \cite{Balasubramanian:2005zx} models. Following the conjecture that dS space can not be obtained in string theory as a matter of principle \cite{Danielsson:2018ztv, Obied:2018sgi}, these models have been the subject of renewed scrutiny. Arguably, the most pressing unresolved issues are the singular bulk problem of KKLT and the related control issues of the LVS \cite{Carta:2019rhx, Gao:2020xqh, Junghans:2022exo, Gao:2022fdi, Junghans:2022kxg, Moritz:2025bsi}.\footnote{
For a selection of other work discussing problems with, as well as arguments in favor of, stringy dS models see \cite{Bena:2009xk, McGuirk:2009xx,  
Kallosh:2015nia,
Polchinski:2015bea,
Bena:2018fqc, Blaback:2019ucp, Blumenhagen:2019qcg, Bena:2020xrh, Lust:2022xoq}. For recent work towards realising stringy dS more explicitly, see e.g.~\cite{Crino:2020qwk, Crino:2022zjk, McAllister:2024lnt}.
}
Both problems involve the anti-D3-brane uplift and become significantly worse if $\alpha'$ corrections in the warped throat region are taken into account \cite{Hebecker:2022zme, Schreyer:2022len}.
While one may hope that other uplifting mechanisms can replace the anti-D3 uplift, this is also not straightforward.\footnote{
See 
\cite{Cicoli:2013cha,Cicoli:2015ylx} for examples and \cite{Balasubramanian:2004uy,Westphal:2005yz,Antoniadis:2018hqy,Antoniadis:2019rkh} for dS proposals without an uplift.
} 
In particular, the complex structure $F$-term uplift \cite{Saltman:2004sn, Denef:2004cf, Gallego:2017dvd, Honma:2019gzp, Hebecker:2020ejb, Dubey:2023dvu, Krippendorf:2023idy, Lanza:2024uis}, which has the potential to be studied very explicitly, has recently been shown to have fundamental control issues in the KKLT context due to warping \cite{Hebecker:2025tui}. 

Against this background the LVS, possibly combined with a complex structure
$F$-term uplift, appears to be a particularly promising route towards controlled de Sitter constructions. This expectation rests on having sufficient control over the underlying, pre-uplift LVS AdS vacuum. The latter is famously generated by the interplay between
non-perturbative effects and perturbative corrections, as quantified within the framework of four-dimensional $\mathcal N=1$ supergravity \cite{Balasubramanian:2005zx}. To complete our understanding, it is important to ask
whether the relevant non-perturbative scalar potential can also be recovered directly from the ten-dimensional effective theory, without using the four-dimensional supergravity structure as an input. The main point of this paper is that,
within our present understanding, the result of such a direct 10d derivation seems to differ from the familiar 4d result already at the leading parametric level.

The motivation for this comparison extends beyond the phenomenological applications of the LVS. It concerns how non-perturbative dynamics localised near a divisor is encoded in a 4d EFT when the overall volume becomes parametrically large. Under the locality assumptions used here, the 10d estimates and the standard 4d supergravity description predict different volume scalings if one retains the tree-level identification of physical divisor volumes with the real parts of chiral coordinates. Reconciling them therefore requires additional input already at leading parametric order.
This could take the form of logarithmic corrections to that identification, or reflect a sensitivity of the non-perturbative sector to the bulk geometry that our local estimates do not capture.
A direct microscopic calculation, e.g., along the lines of \cite{Sen:2021qdk,Sen:2021tpp,Sen:2021jbr,Alexandrov:2021shf,Alexandrov:2021dyl,Alexandrov:2022mmy} or of \cite{Kaufmann:2026mha}, could potentially lead to progress.

A closely related issue has appeared before, although in a somewhat different guise. The gauge-threshold analyses of local D$7$-brane sectors in \cite{Conlon:2009kt,Conlon:2009xf} showed that local gauge couplings can retain logarithmic sensitivity to bulk scales, and that consistency with the Kaplunovsky--Louis formula \cite{Kaplunovsky:1993rd,Kaplunovsky:1994fg} may be encoded in one-loop redefinitions of the holomorphic variables. Conlon and Pedro \cite{Conlon:2010ji} then analysed the implications of such K\"ahler coordinate redefinitions for moduli stabilisation in the LVS. In particular, for a small blow-up cycle supporting gaugino condensation, a logarithmic shift can modify the relation between the physical cycle volume and the holomorphic variable entering the non-perturbative superpotential.\footnote{For related discussions of logarithmic corrections, K\"ahler coordinate redefinitions and perturbative LVS-type stabilisation in nearby settings, see e.g.~\cite{Antoniadis:1998ax,
Cicoli:2012fh, deAlwis:2012bm, Grimm:2017pid,Weissenbacher:2019mef,Weissenbacher:2020cyf, Antoniadis:2018hqy,Antoniadis:2019rkh,Leontaris:2022rzj,Klaewer:2020lfg, Cicoli:2024bwq, Hai:2025wvs,Chakraborty:2025wqn}. We are not aware, however, of a derivation of precisely the correction which would be needed in the present setup to reconcile a strictly local ten-dimensional scaling estimate with the standard LVS potential.} Thus the previous lesson was relatively mild: the effect was treated as a possible coordinate correction, with model-dependent coefficient, rather than as evidence for a parametric failure of the usual LVS description.

We want to make a stronger and more general point: both in the instanton and in the gaugino-condensate version, a naive 10d EFT derivation of the three-term LVS scalar potential appears to lead to a mismatch in two of the three terms. The discussion of \cite{Conlon:2010ji} was tied primarily to D7-brane gauge thresholds and hence to the gaugino-condensation realisation of the LVS. The point we wish to emphasise is that the same type of tension also appears in the E3-instanton realisation, where there is no immediate four-dimensional gauge-threshold interpretation. Thus, the instanton realisation is not automatically insulated from the issue, but requires an independent ten-dimensional account of how the relevant holomorphic K\"ahler coordinate, or equivalently the normalisation of the instanton contribution, arises. This mismatch between the 10d EFT estimate and the standard 4d SUGRA result can be removed if one allows for a logarithmic correction to the relation between the physical small-cycle volume and the corresponding chiral coordinate on the K\"ahler manifold. The coefficient of the logarithm is then not a freely tunable small number: it is fixed by matching the LVS scalings, is of order one, and changes the familiar relation between the small-cycle physical volume and $\log{\cal V}$ by a factor of two. More generally, the logarithmic redefinition represents only one possible way of encoding the additional volume dependence required by the matching. The microscopic origin of the necessary effect remains unclear, and different ways of realising it may have very different implications.
As argued above, a reliable understanding of the LVS AdS vacuum is essential in view of the justified doubts concerning de Sitter constructions in string theory. Thus, there is work to be done.

The rest of the paper is organised as follows: In Sec.~\ref{lvse}, we recall the relevant 4d supergravity description as well as the 10d EFT derivation of the $|W_0|^2/{\cal V}^3$ term in the scalar potential, which is not problematic. Section~\ref{instt} treats the flux-instanton term and the instanton-anti-instanton terms, with conventional scaling $W_0\exp(-a\tau_s)/{\cal V}^2$ and
$\exp(-2a\tau_s)/{\cal V}$ respectively. As already spelled out in the abstract, these scalings are not recovered. The analogous problem arising in the gaugino-condensate version of the LVS is studied in Sec.~\ref{gcon}. Given these results, we discuss possible ways out in Sec.~\ref{wout}: In particular, we develop the suggestion of \cite{Conlon:2010ji} by allowing for log corrections to both the fields and to the functional form of the K\"ahler potential. We also avoid any (in our opinion unjustified) perturbative expansions in the prefactor of the log correction. Finally, we argue in some detail why more work is needed to be certain that the LVS AdS vacuum is understood. We conclude in Sec.~\ref{conc}.

\section{Large volume scalar potentials and a first step towards 10d}\label{lvse}

Let $X$ denote the Calabi--Yau orientifold underlying our compactification, and let $\{\omega_i\}$ be a basis of $H^{1,1}_+(X)$. The K\"ahler form may be expanded as
\begin{equation}
    J = t^i \omega_i\,,
\end{equation}
where we take $t^i$ to measure 2-cycle volumes in the 10d Einstein frame in string units.\footnote{
More precisely, we absorb the $1/g_s^2$ prefactor of the Einstein-Hilbert term in the metric and set $2\pi\sqrt{\alpha'}$ to unity.
}
The Calabi--Yau volume can be written as
\begin{equation} 
    \mathcal V = \frac{1}{3!}\int_X J\wedge J\wedge J = \frac{1}{6}\,\kappa_{ijk}t^i t^j t^k\,,\qquad \kappa_{ijk} = \int_X\omega_i\wedge\omega_j\wedge\omega_k\,.
\end{equation}
Moreover, the Einstein-frame volume of the divisors $\Sigma_i$ dual to $\omega_i$ reads
\begin{equation}
    \tau_i \equiv \text{Vol}_E(\Sigma_i) = \frac{1}{2}\int_{\Sigma_i}J\wedge J = \frac{\partial\mathcal V}{\partial t^i} = \frac{1}{2}\,\kappa_{ijk}t^j t^k\,.
\end{equation}

Our interest is in scalar potentials in the large volume limit. More specifically, we will uncover problems in the understanding of localised non-perturbative effects associated with 4-cycles which can stay finite as ${\cal V}\to \infty$. The prototypical example of this is the LVS setting, but the issue is much broader, as will become apparent below. For concreteness, we will nevertheless focus on the LVS constructions \cite{Balasubramanian:2005zx} based on a two-modulus Calabi--Yau orientifold with Swiss-cheese structure, such that the volume can be expressed as 
\begin{equation}
    \mathcal V=\frac{1}{\lambda}\left(\tau_{b}^{3/2}-\tau_{s}^{3/2}\right)\,.
\end{equation}
The 4d supergravity description is obtained in terms of 
\begin{equation}
    T_{b,\ho}=\tau_{b,\ho}+\I \rho_b\,,\qquad
    T_{s,\ho}=\tau_{s,\ho}+\I \rho_s,
\end{equation}
referred to as large- and small-cycle moduli, respectively. At tree level, the supergravity variables are identified as 
\begin{equation}
    \tau_{i,\ho} \equiv \tau_{i}\, , \qquad \rho_i \equiv C_4\text{-axions}
\end{equation}

Including $\alpha'$-corrections to the K\"ahler potential and non-perturbative effects in the holomorphic superpotential, the model is characterised by 
\begin{equation}
    K=-2\log\left(\mathcal V+\frac{\xi}{2g_s^{3/2}}\right),\qquad W=W_0+A \mathrm{e}^{-aT_{s,\ho}}\, .
\end{equation}
Employing standard 4d $\mathcal{N}=1$ supergravity expressions, the large volume expansion of the scalar potential schematically reads
\begin{equation}\label{eq:standard-LVS-Potential}
	V_{\text{LVS}} \;  \sim \; \frac{a^2A^2\sqrt{\tau_{s,\ho}}\, \mathrm{e}^{-2a\tau_{s,\ho}}}{\mathcal V}-\frac{aAW_0\tau_{s,\ho}\,  \mathrm{e}^{-a\tau_{s,\ho}}}{\mathcal V^2}+\frac{\xi W_0^2}{g_s^{3/2}\mathcal V^3}\, .
\end{equation}

In what follows, we want to verify whether this scalar potential can be understood at the parametric level from the compactification of the 10d effective theory. 
Let us start by considering the last term in \eqref{eq:standard-LVS-Potential}, following \cite{Conlon:2005ki,Cicoli:2021rub,Liu:2022bfg}. Its 10d origin is clear: it follows from the dimensional reduction of the higher-derivative $(\alpha')^3$ terms; the relevant pieces in the 10d action schematically read,
\begin{equation}
   S_{(\alpha')^3}\supset \int \dif^{10}x\sqrt{-g}\left(R^4+R^3|G_3|^2+R^3G_3^2+R^3\overline{G}_3^2+\cdots\right)\, ,
\end{equation}
where the different powers $R^n$ denote the relevant contractions of the Riemann tensor. The purely gravitational $R^4$ coupling
gives the BBHL correction to the four-dimensional K\"ahler potential \cite{Antoniadis:1997eg,Becker:2002nn,Antoniadis:2003sw}.\footnote{The original derivations of the BBHL correction used the tree-level
$R^4$ coupling. The corresponding eight-derivative sector with
$G_3$ insertions was organised in an $\mathrm{SL}(2,\mathbb Z)$-covariant form
in \cite{Liu:2022bfg}, thereby including the perturbative and D$(-1)$-instanton
completion of the relevant modular coefficients. The
same ten-dimensional $R^4$ coupling also induces higher-derivative
interactions in 4d, including the $F^4$ corrections discussed in
\cite{Ciupke:2015msa}.}
In a background with three-form flux, the terms
with two insertions of $G_3$ give the ten-dimensional representation of
the corresponding $W_0^2$ contribution to the scalar potential. More
precisely, both the neutral coupling $R^3 |G_3|^2$ and the
$\mathrm{U}(1)_R$-charged couplings $R^3G_3^2+R^3\overline{G}_3^2$
contribute to this term after evaluating the action on the flux
background.\footnote{The role of the charged terms was already noted in
\cite{Becker:2002nn}. The relevant five-point couplings were computed
from string amplitudes in \cite{Liu:2019ses} and were further organised
using duality and M-theory methods in \cite{Liu:2022bfg}.}
The topological dependence follows from the fact that the internal
curvature contraction reduces to the six-dimensional Euler density\footnote{The $\alpha'^3$ correction also receives an additional order-one contribution sensitive to the orientifold/7-brane sector (see \cite{Minasian:2015bxa}).},
\begin{equation}
    \int_X \dif^6y\sqrt{g_6}\,R_6^3\sim \chi(X)\sim \xi \,.
\end{equation}
Moreover, the flux density may be expressed as
\begin{equation}
   |G_3|^2 \sim\frac{|W_0|^2}{\mathcal V}\,,\qquad\mbox{with}\qquad
    W_0=\int_XG_3\wedge\Omega
    \label{eq:G3_norm}
\end{equation}
the flux-induced superpotential. This volume scaling follows simply from the three inverse metric factors in the contraction, as described e.g.~in~\cite{Conlon:2005ki} (see also our App.~\ref{app:g3s}). As a result, ignoring $g_s$ factors for brevity, the dimensional reduction of $R^3|G_3|^2$ parametrically gives \cite{Conlon:2005ki,Cicoli:2021rub,Liu:2022bfg}
\begin{equation}
\xi\times \frac{|W_0|^2}{\mathcal V}\times \frac{1}{{\cal V}^2}\,,
\end{equation}
where the last factor comes from Weyl rescaling to the 4d Einstein frame. Thus, we straightforwardly reproduce the perturbative contribution to \eqref{eq:standard-LVS-Potential}.

\section{Volume scaling of the E3 instanton effects}\label{instt}

In what follows, we will investigate whether the non-perturbative contributions appearing in the scalar potential \eqref{eq:standard-LVS-Potential} admit a similar 10d description. 
In this section, we assume the non-perturbative effect stabilising the small 4-cycle to arise from a Euclidean D3-brane instanton wrapping the rigid divisor $\Sigma_s$. 

\subsection{Some instanton basics}
Let us start by recalling more generally how the instanton contribution to the partition function ${\cal Z}$ affects the vacuum energy. For simplicity, we consider a purely bosonic theory; the crucial role played by fermionic zero modes will be discussed at the end.

To begin with, let us distinguish the light scalar fields $\varphi^a$ of our theory, whose potential we wish to determine, from the remaining fields and fluctuations, collectively denoted by $\Phi$. For a given background value of the light fields, the Euclidean path integral reads:
\begin{equation}
    \mathcal{Z}[\varphi] \quad = \quad \int \mathcal{D}\Phi \,\mathrm{e}^{-S_E[\varphi,\Phi]}\, .
\end{equation}
In particular, one can define an effective Euclidean action\footnote{In what follows, we will consider only the classical configurations of the light fields $\varphi$ and ignore their quantum fluctuations.}
\begin{equation}
    S_{\rm eff}[\varphi] = - \log \mathcal{Z}[\varphi]\, .
\end{equation}
For constant values of the fields, the derivative terms in the action vanish and one finds
\begin{equation}
    S_{\rm eff}[\varphi] = ({\rm Vol}_4) V_{\rm eff}\, ,
\end{equation}
or, equivalently,
\begin{equation}\label{eq:V_effective_light_fields}
    V_{\rm eff}(\varphi) = - \frac{1}{{\rm Vol}_4} \log \mathcal{Z}[\varphi]\, .
\end{equation}
The path integral admits a semi-classical expansion into distinct instanton sectors
\begin{equation}
    \mathcal{Z}[\varphi] =\mathcal{Z}_0[\varphi] +
    \mathcal{Z}_{I}[\varphi] +
    \mathcal{Z}_{I\bar{I}}[\varphi] +\cdots \, ,
\end{equation}
where $\mathcal{Z}_I$ denotes the one-instanton contribution and $\mathcal{Z}_{I\bar{I}}$ the instanton/anti-instanton contribution.
Combined with equation \eqref{eq:V_effective_light_fields}, we immediately see that
\begin{equation}
    V_{\rm eff}(\varphi) \simeq -\frac{\log\mathcal{Z}_0}{{\rm Vol}_4} -\frac{1}{{\rm Vol}_4} \frac{\mathcal{Z}_{I}[\varphi]}{\mathcal{Z}_0[\varphi]} +\cdots\,  .
\end{equation}
Therefore, if the one-instanton effect has the form
\begin{equation}
    \frac{\mathcal{Z}_{I}}{\mathcal{Z}_0} \;  \sim \;  {\rm Vol}_4\,\kappa(\varphi)\,\mathrm{e}^{-S_{\text{inst}}(\varphi)}\, ,
\end{equation}
its contribution to the effective potential $V_{\rm eff}(\varphi)$ reads
\begin{equation}
     V_{I}(\varphi) \;  \sim \;  -\kappa(\varphi)\,\mathrm{e}^{-S_{\text{inst}}(\varphi)}\,  .
\end{equation}
To conclude, let us recall that, in a fully-fledged supersymmetric theory, the inclusion of fermions does not modify this general structure. Crucially, however, as will be clear below, an instanton sector contributes to the bosonic effective potential only if its fermionic zero modes are lifted by interaction terms; otherwise, the corresponding Grassmann integrations make its contribution to the vacuum functional vanish.

\subsection{The instanton-flux term}

\begin{figure}[t!]
\centering
\includegraphics[width=\textwidth]{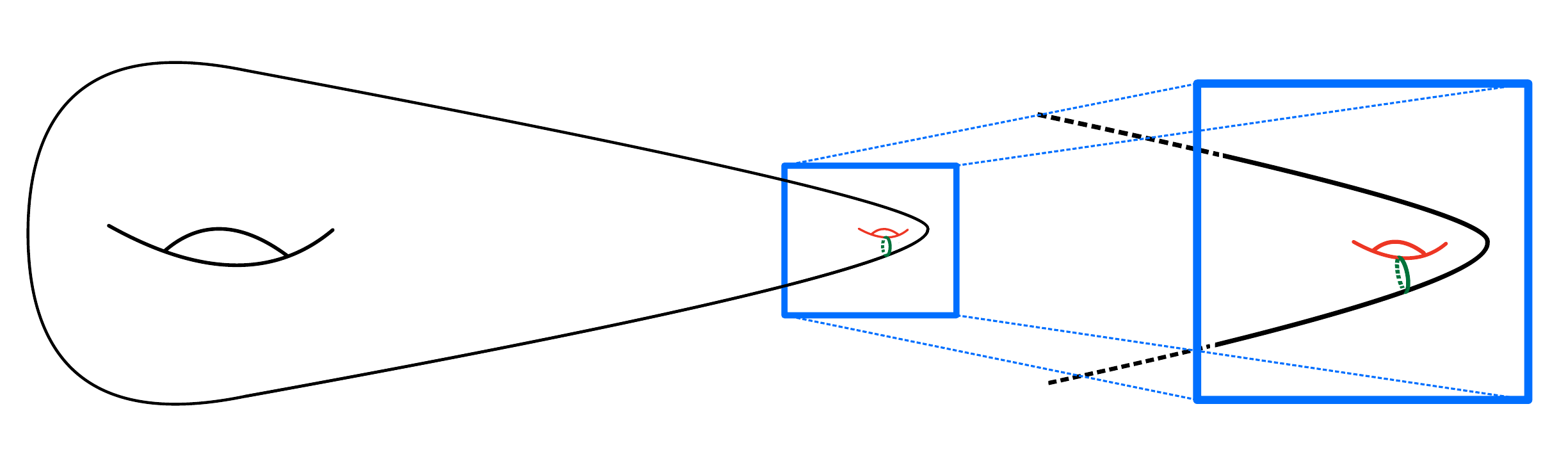}
\caption{Schematic illustration of an LVS-type geometry, where $\tau_s\ll {\cal V}$ and the volume can, in principle, be taken to infinity.
The region in the blue box, shown enlarged on the right, can at low energies be characterised by a 4d `local QFT'.}
\label{fig:local_E3_QFT}
\end{figure}

We focus first on the single--exponential term in the supergravity scalar potential, in the large volume expansion:
\begin{equation}
    V_{\text{lin}}^{\text{LVS}} \;  \sim \; -\frac{aAW_0\,\tau_{s,\ho}\, \mathrm{e}^{-a\tau_{s,\ho}}}{\mathcal V^2}\, .
    \label{eq:E3-LVS-linear-required}
\end{equation}
For our purposes, the relevant path integral is that of the quantum field theory living in the region of the Calabi--Yau where the small 4--cycle is localised (cf.~Fig.~\ref{fig:local_E3_QFT}). The dynamical degrees of freedom of this `local 4d QFT' are therefore the instantons and the small-cycle modulus $\tau_s$ with its superpartners.\footnote{
One could consider including local field fluctuations, e.g.~the fluctuations of the metric near an E3-instanton, which appear to be natural partners of the fluctuations of the instanton itself. On the one hand, these are too heavy for our low-energy QFT, but on the other hand they are hard to separate from the bulk KK-modes, some of which become light when ${\cal V}\to\infty$. We will for the moment assume that integrating out such local KK modes does not break the `local QFT' picture we have argued for.
} 
We assume that bulk fields, such as the axio-dilaton or the 3-form fluxes, enter only as background expectation values, which determine the couplings of the local theory. In particular, we assume that the relevant local quantities admit a well-defined non-compact limit, obtained by sending ${\cal V}$ to infinity while keeping $\tau_s$ and local background fields fixed.

Thus, while a careful treatment of the path integral over fluctuations around the different instanton backgrounds is needed to determine numerical coefficients, locality implies that these coefficients should tend to finite constants as $\mathcal{V}\to\infty$. In other words, dependence on the global bulk volume should appear only through corrections that vanish in the $\text{large-}\mathcal{V}$ limit.
Under these assumptions, we may disregard normalisation issues of the instanton pre-factor in the metric frame where the volume decouples, as just explained. 

In this framework, the one-instanton contribution reads
\begin{equation}
    \mathcal{Z}_{I}[g_{\mu\nu},G_3,\cdots] \;  \sim \; 
    \int
    \mathcal D\Phi\,
    \exp\!\left(-S_{E3}[T_s, \Psi,\Phi]\right)\, ,
    \label{eq:E3-1I-path integral}
\end{equation}
where $T_s=\tau_s+\I\rho_s$, with $\tau_s$ denoting the 10d Einstein frame volume of the four-cycle wrapped by the Euclidean D3-brane, $\Psi$ stands for the fermionic superpartners of $T_s$ and $\Phi$ collectively describes the quantum fluctuations around the instanton solution (both bosonic and fermionic). We suppress the dependence on bulk fields to avoid cluttering the notation.

The path integral factorises into an integral over zero modes and a determinant factor associated with massive fluctuations. 
For the present purpose, we consider an $O(1)$ instanton wrapping a rigid divisor. In this case, the orientifold projection removes half of the universal fermionic zero modes and leaves only two zero modes $\theta^\alpha$ related to the 4d superspace coordinates.\footnote{For a comprehensive review, see \cite{Blumenhagen:2009qh,Ibanez:2012zz}. See also
\cite{Bianchi:2011qh} for a detailed analysis of E3-brane fermionic zero modes. Note that, in the standard supergravity description, the presence of precisely two unlifted fermionic zero modes is the criterion for a contribution to the 4d superpotential.}
The partition function can therefore be written as
\begin{equation}
    \mathcal{Z}_{I} \;  \sim \; 
    \mathcal{A}_{\text{fluct}} \times \int \dif^4 x_I\,\dif^2\theta\; 
    \exp\left[-S_{E3}\right]\, .
    \label{eq:E3-one-instanton-zero-modes}
\end{equation}
Here $x^\mu_I$ are the bosonic 4d instanton coordinates and $\mathcal{A}_{\text{fluct}}$ is obtained by integrating over fluctuations around the classical instanton solution. The instanton suppression factor is
\begin{equation}
    \mathrm{e}^{-S^{(0)}_{E3}}\equiv\mathrm{e}^{-a T_s}\,.
\end{equation}
In the absence of background fluxes $G_3$, no interaction term is present to saturate the fermionic measure, and the contribution to the vacuum energy vanishes identically:
\begin{equation}
    \mathcal{Z}_{I}\big|_{G_3 \,=\, 0} \;  \sim \; \mathcal{A}_{\text{fluct}}\int \dif^4x_I \int \dif^2\theta\,\mathrm{e}^{-S_{E3}}=0\,.
\end{equation}

If $G_3$ is non-zero, we have to include the appropriate piece of the fermionic D3-brane action in our analysis. It takes the schematic form \cite{Grana:2002tu,Camara:2004jj,Tripathy:2005hv,Martucci:2005rb,Bianchi:2012kt}
\begin{equation}
S_{E3} \supset\int_{\Sigma_s}\dif^4y\sqrt{g_{\Sigma_s}} \; G_{mnp} \overline{\lambda}\Gamma^{mnp} \lambda\,,
    \label{eq:E3-flux-fermion-coupling}
\end{equation}
where $\Gamma^{\mu\nu\rho} \equiv e^{\mu}_a e^{\nu}_be^{\rho}_c \, \Gamma^{abc}$ in terms of the vielbein $e^{\mu}_a$ and $\lambda$ denotes the E3-brane fermion. Expanding the E3-brane fermion in the universal zero-mode sector as
\begin{equation}
    \lambda = \theta^\alpha\psi_{\theta,\alpha} + \ldots
\end{equation}
and normalising the wavefunctions $\psi_{\theta,\alpha}$ with respect
to the fermionic kinetic inner product gives
\begin{equation}
    S_{E3}
    \supset
    m_{\theta\theta}\,\theta\theta\,,
    \qquad
    m_{\theta\theta}
    \;\sim\;
   \int_{\Sigma_s}\dif^4y\sqrt{g_{\Sigma_s}} \;
   G_{mnp}\,\bar\psi_\theta\Gamma^{mnp}\psi_\theta\,.
    \label{eq:E3-theta-theta-flux}
\end{equation}
Here the flux-dependent interaction is treated as an insertion in the semi-classical expansion around the fluxless E3 saddle.

For an $\mathrm{O}(1)$ E3-instanton in an ISD GKP-type background, the flux-selection rules imply that the universal $\theta\theta$ bilinear couples to the $(0,3)$ component of $G_3$. By contrast, primitive $(2,1)$ flux couples to other components of the E3-brane fermion and does not lift the two universal $\theta$ modes \cite{Tripathy:2005hv,Blumenhagen:2009qh}. Hence, defining
\begin{equation}
    B_\theta^{mnp}   \equiv \bar\psi_\theta\Gamma^{mnp}\psi_\theta\,,
\end{equation}
the modulus of the integrand in \eqref{eq:E3-theta-theta-flux} obeys 
\begin{equation}
    \left|G_{mnp}\,\bar\psi_\theta\Gamma^{mnp}\psi_\theta\right| =\left|G^{(0,3)}_{mnp}B_\theta^{mnp}\right| \leq |G_{(0,3)}|\,|B_\theta|\,.
\end{equation}
After integration over the small divisor, we may parametrise the result as
\begin{equation}
    |m_{\theta\theta}|   =    |c_\theta(\tau_s)|\,|G_{(0,3)}|\,,
\end{equation}
where $c_\theta(\tau_s)$ collects numerical factors and dependence on the local geometry and zero-mode profile. As shown in Appendix~\ref{app:g3s}, we have
\begin{equation}
    |G_{(0,3)}|    =  \frac{|W_0|}{\sqrt{\mathcal V}}\,
    \label{eq:G03loc}
\end{equation}
so that, with $c_\theta(\tau_s)$ understood as a generally complex coefficient, we obtain
\begin{equation}
    m_{\theta\theta}    =     c_\theta(\tau_s)\,    \frac{W_0}{\sqrt{\mathcal V}}\,.
\end{equation}

This conclusion assumes that the canonically normalised universal zero-mode wavefunctions admit a well-defined $\mathcal{V}\rightarrow\infty$ limit at fixed $\tau_s$, so that their normalisation introduces no additional power of the bulk volume.
This issue is closely related to, but distinct from, the volume-dependent D7-gaugino normalisation discussed in Sec.~\ref{sec:Vgaugino}. In the E3-instanton calculation, the relevant question concerns the consistent normalisation of the universal fermionic collective coordinates, their integration measure and the flux-induced $\theta\theta$ insertion.

The flux bilinear then saturates the two universal Grassmann integrations
\begin{equation}
    \int\dif^2\theta\,
    \left(m_{\theta\theta}\theta\theta\right)\, \mathrm{e}^{-S_{E3}^{(0)}}
    =
    m_{\theta\theta}\,
    \mathrm{e}^{-aT_s}\,.
\end{equation}
The resulting one-instanton contribution
therefore takes the parametric form
\begin{equation}
    \frac{\mathcal{Z}_I}{\mathcal{Z}_0}
    \simeq
    \mathrm{Vol}_4\times
    C_I(\tau_s)\,
    |G_{(0,3)}|\,
    \mathrm{e}^{-aT_s}
    \sim
    \mathrm{Vol}_4\times
    C_I(\tau_s)\,
    \frac{W_0}{\sqrt{\mathcal{V}}}\,
    \mathrm{e}^{-aT_s}\,.
\end{equation}
Here $C_I(\tau_s)$ collects all numerical pre-factors and $\tau_{s}$-dependence coming from the fluctuations around the background and the interaction term, while $\mathrm{Vol}_4$ arises from integration over the four translational zero modes $x_I^\mu$. The anti-instanton supplies the complex-conjugate contribution. Since we are only interested in the leading dependence on $\tau_s$, we henceforth suppress the sub-exponential $\tau_s$-dependence of $C_I (\tau_s)$ and simply write $C_I$. 

The leading dilute-gas contribution to the vacuum functional is then schematically given by
\begin{equation}
    \mathcal{Z} \simeq\mathcal{Z}_0 \left[
        1 + {\rm Vol}_4  \; |{G_{(0,3)}}|\left(C_I \;  \mathrm{e}^{-a T_s} + \overline{C_I} \;  \; \mathrm{e}^{-a \overline{T_s}} \right)+ \cdots \right].
\end{equation}
The corresponding contribution to the scalar potential reads 
\begin{equation}
    V^{\text{BD}}_{I} \simeq |{G_{(0,3)}}|\left(C_I \;  \mathrm{e}^{-a T_s} + \overline{C_I} \;  \; \mathrm{e}^{-a \overline{T_s}} \right)\,.
\end{equation}

Therefore, employing \eqref{eq:G03loc}, in the Brans-Dicke frame, the one-instanton contribution to the scalar potential scales as 
\begin{equation}
    V^{\text{BD}}_{I} \,\simeq \,C_I \,\frac{W_0 }{\sqrt{\mathcal{V}}}\,  \mathrm{e}^{-a\tau_{s}} + {\rm c.c}\, .
\end{equation}
The Weyl rescaling to the 4d Einstein frame contributes the familiar factor $\mathcal V^{-2}$, giving the scalar potential
\begin{equation}
    V_{I} \, \simeq \, C_I\,\frac{ W_0 }{\mathcal{V}^{5/2} } \,  \mathrm{e}^{-a\tau_{s}} + {\rm c.c}\,.
    \label{eq:E3-linear-naive}
\end{equation}
Thus, if one identifies the supergravity field with the physical divisor volume,
\begin{equation}
    \tau_{s,\ho}=\tau_{s}\, ,
\end{equation}
the 10d scaling estimate fails to reproduce the volume dependence of the single-exponential term predicted by 4d supergravity.

\subsection{The instanton/anti-instanton term}

We now turn to the term in \eqref{eq:standard-LVS-Potential} which is quadratic in the instanton suppression, again in the large volume limit:
\begin{equation}
    V_{\text{quad}}^{\text{LVS}} \;  \sim \; \frac{a^2A^2\sqrt{\tau_{s,\ho}}\, \mathrm{e}^{-2a\tau_{s,\ho}}}{\mathcal V}\,.
\end{equation}
As argued in \cite{Dine:1987bq} in the context of world-sheet instantons, such terms in the scalar potential are naturally interpreted as instanton/anti--instanton effects. In the present case, suppressing numerical coefficients, the relevant contribution to the partition function can be written as
\begin{equation}
\begin{aligned}
        \mathcal Z_{I\bar I} &\sim \int \dif^4x_I \, \dif^4x_{\overline{I}} \,\dif^2\theta_I \,  \dif^2\theta_{\overline{I}}\, \exp\left[-S^{(0)}_{E3}-S^{(0)}_{\overline{E3}}-S_{\text{int}}(I,\bar I,\Phi_\text{bulk})\right]\\
        & \sim \mathrm{e}^{-aT_s-a\overline{T_s}} \int \dif^4x_I \, \dif^4x_{\overline{I}} \,\dif^2\theta_I \,  \dif^2\theta_{\overline{I}} \, \mathcal{F}_{\text{int}}\left( |x_I-x_{\overline{I}}|,\theta_I,\theta_{\overline{I}},\Phi_\text{bulk}\right)\\
        & \sim {\rm Vol}_4 \times \mathrm{e}^{-2a\tau_{s}}  \int r^3 \dif r \,\dif^2\theta_I \,  \dif^2\theta_{\overline{I}} \, \mathcal{F}_{\text{int}}\left( r,\theta_I,\theta_{\overline{I}},\Phi_\text{bulk}\right)
\end{aligned}
\label{eq:IIbar-sector}
\end{equation}
where we made use of the fact that
\begin{equation}
    S^{(0)}_{E3} = aT_s, \qquad S^{(0)}_{\overline{E3}}=a\bar T_s,
\end{equation}
and we collectively denoted by $\mathcal{F}_{\text{int}} \equiv \exp\left[-S_{\text{int}}(I,\bar I,\Phi_\text{bulk})\right]$ interactions within each instanton sector, mutual interactions between the instanton and anti-instanton, and their dependence on relative collective coordinates and local bulk fields $\Phi_{\text{bulk}}$. In particular, the mutual interactions are expected generically to lift the fermionic zero modes of each instanton, making it possible for this sector to contribute to the scalar potential.

An explicit computation of $\mathcal F_{\text{int}}$ is beyond the scope of this work. Such a calculation would require control over the correlated instanton/anti--instanton background and over the corresponding fluctuation determinants. 
However, for our purposes, only its parametric dependence on the bulk volume $\mathcal V$ is relevant. The instanton and anti-instanton both wrap the same small cycle $\Sigma_s$, and the dynamics that determines their local interaction is described by the local QFT living in the small 4--cycle region.
As argued before, under our locality assumptions, the general field theory expectation is that no leading-order $\mathcal V$-dependence arises.

Since this is a crucial point, let us argue the expected volume-independence more carefully: At every value of $r$, our basic configuration is a pair of E3 branes, wrapped on the small cycle in an identical, volume-minimising way. They are separated by a distance $r$ in the non-compact directions. The crucial $\theta$-dependence arises only for small $r$ and is due to the backreaction of each brane on the background, which then affects the fermionic action of the other brane. Thus, the effect comes from the propagation of 10d fields from one brane to the other, with both branes being localised in the small-cycle region and separated by a small amount in the non-compact directions. We expect the relevant separation scale, i.e.,~the typical value of $r$, to be the small cycle size or smaller.\footnote{
The logic behind this expectation is that, once $r\gg \tau_s^{1/4}$, the instantons `see' each other through the propagation of massless bulk fields in some effective 10d geometry. Concretely, the relevant geometry is the product of a 6d cone $C$ (depending on the precise type of singularity arising at $\tau_s\to 0$) and flat 4d space: $C\times \mathbb{R}^4$. We need to know the propagator between two points at the tip of the cone, separated by $r$ in the flat directions. The natural expectation is $\sim 1/r^8$, based on the fact that the relevant geometry, though not $\mathbb{R}^{10}$, is properly 10-dimensional as long as the base of the cone does not strongly prefer only a subset of the compact dimensions. We conclude that the radial integral appearing in the partition function is of type $r^3\,\mathrm{d}r\,/r^8$ and hence dominated by small $r$, as claimed.

We note that there is a caveat to the above logic, which was pointed out to us by Joe Conlon: For 4-cycles or 2-cycles, one may define Wilson-line-type fields, such as $\oint C_2$. One may think of those as propagating along the cycle which is dual to the one used in the integration above. Depending on the topology of cycles in the bulk and the sourcing of the relevant RR fields by the instanton, it is then conceivable that an effective 2d propagator and a log divergence arise. Demonstrating or disproving this goes beyond the scope of this work.
}
From this picture it is natural to expect that a non-zero effect, if present at all, will not be enhanced if the volume is taken to infinity: The propagators in question are simply not sensitive at leading order to the `bulk' geometry very far away from the small-cycle region. 

Therefore, repeating the analysis outlined above for the single instanton case, the instanton/anti--instanton contribution to the scalar potential reads
\begin{equation}
    V^{\rm BD}_{I\bar I} \simeq C_{I\bar I}\, \mathrm{e}^{-2a\tau_{s}}\,.
    \label{eq:IA_V_contribution_BD}
\end{equation}
Crucially, while $C_{I\bar I}$ is not determined by our analysis, it should not depend on ${\cal V}$.
In the 4d Einstein frame, one then finds
\begin{equation}
    V_{I\bar I} \simeq C_{I\bar I}\, \frac{\mathrm{e}^{-2a\tau_{s}}}{\mathcal{V}^2}\, .
    \label{eq:IA_V_contribution_E}
\end{equation}
Thus, also in this case, the volume scaling obtained from the 4d analysis is not reproduced if $\tau_{s,\ho}=\tau_{s}$.

\section{Volume scaling of the gaugino condensate}\label{gcon}

We now turn to the second possibility of realising the non-perturbative effect used in moduli stabilisation: the D7-brane gaugino condensate. The 10d EFT perspective of the volume scaling of this effect has already been discussed in \cite{Conlon:2009xf,Conlon:2009kt,Conlon:2010ji}.

Here, we first show that the local EFT analysis leads, as in the E3-instanton case, to the wrong bulk-volume scaling of the non-perturbative scalar potential. We then use the Kaplunovsky--Louis formula to show that agreement with four-dimensional supergravity would require an effective UV scale enhanced relative to the naive local cutoff. Finally, we review \cite{Hamada:2021ryq}, where the standard scaling is recovered, and point out that their matching implicitly requires a bulk-volume enhancement of the higher-dimensional gaugino bilinear, which appears unnatural from the perspective of a local 10d EFT.

\subsection{EFT expectations for the scalar potential}

The simplest realisation of the effect arises if the small cycle is wrapped by a stack of four D7 branes on top of an O7 plane, such that the D7-tadpole is cancelled locally. This gives rise to a 4d $\mathrm{SO}(8)$ $ {\cal N}=1$ super Yang-Mills theory with inverse coupling identified at tree level as
\begin{equation}\label{eq:YMtree}
    \frac{4\pi}{g^2} = \tau_{s}\, .
\end{equation}
Assuming this matching to be valid at some UV scale $\Lambda_{\rm UV}$, the coupling then runs according to
\begin{equation}
    \frac{4\pi}{g^2(\mu)} = \tau_{s} + \frac{\beta_0}{4\pi} \log\left(\frac{\Lambda_{\text{UV}}^2}{\mu^2}\right)\, ,
    \label{eq:D7_log_running}
\end{equation}
where we introduced 
\begin{equation}
    \beta_0 = \sum_R n_R T(R) - 3T(G)
\end{equation}
and $T(G)$ is the Dynkin index of the adjoint of $G$, see also~\eqref{eq:gauge_generators_conventions}.\footnote{Recall that for $G =\mathrm{SO}(N)$, we have $T(\mathrm{SO}(N)) = N-2$.}

The running described in \eqref{eq:D7_log_running} identifies a strong-coupling scale via dimensional transmutation
\begin{equation}
    \Lambda^3_{\text{strong}} = \Lambda^3_{\text{UV}}\, \exp\left(\frac{6\pi}{\beta_0}\tau_{s}\right) \equiv \Lambda^3_{\text{UV}}\,  \mathrm{e}^{-a\tau_{s}}\,.
\end{equation}
The naive expectation is that $\Lambda_{\rm UV}$ is the scale below which our theory effectively becomes 4d, i.e.,~the KK-scale associated to the small-cycle\footnote{Note that in our conventions $M_s\sim 1$, but we keep it explicit in the present formulae for conceptual clarity.}
\begin{equation}
    \Lambda_{\rm UV}\sim M_{KK, \, \tau_s} \sim \frac{M_s}{\tau^{1/4}_{s}} \, .
\end{equation}
Since we are interested in the parametric dependence on the bulk volume $\mathcal{V}$, we may set $M_{\rm KK,\,\tau_s}\sim M_s$, discarding powers of $\tau_s$. On dimensional grounds, the gaugino condensate induced at the strong coupling scale of the local gauge theory  then reads
\begin{equation}
    \langle \lambda\lambda\rangle_{\text{loc}}
    \sim \Lambda_{\rm strong}^3\sim M_s^3 
    \mathrm{e}^{-a\tau_{s}}\, .
\end{equation}
Below $\Lambda_{\rm strong}$, the $\mathrm{SO}(8)$ gauge theory confines and develops a mass gap. We can therefore integrate it out and focus on the modification of the scalar potential of the effective theory of moduli.
Pure ${\cal N}=1$ Super Yang--Mills does not break supersymmetry and hence, by itself, does not generate a vacuum energy \cite{Witten:1982df}. In particular, no scalar potential term $\sim \Lambda_{\rm strong}^{4}$ arises. Rather, the strongly coupled gauge sector affects the low-energy theory through the expectation values of operators appearing in its couplings to the other moduli. 

Below the confinement scale the gaugino condensate is encoded in a non-dynamical source term
\begin{equation}
    {\cal J}_{2} \equiv \left\langle \lambda\lambda\right\rangle_{\rm loc} \sim \Lambda_{\rm strong}^{3} \sim M_s^{3}\mathrm{e}^{-a\tau_s}\,,
    \label{eq:gaugino_bilinear_source}
\end{equation}
which may appear in the low-energy EFT. In particular, in analogy to \eqref{eq:E3-flux-fermion-coupling}, the D7 branes on the small-cycle possess an action term coupling the gaugino-bilinear to the flux \cite{Camara:2004jj}. This induces a contribution to the 4d action
\begin{equation}
    S_{G_3\lambda^2}\;  \sim \; \int \dif^4x\,\sqrt{-g^{\rm BD}_4}\, 
    \tau_s\,G_{(0,3)}\, {\cal J}_2 
    ,
\label{eq:G3_J2_coupling}
\end{equation}
where, as in the instanton case, $G_{(0,3)}$ denotes the (0,3)-component of the bulk three-form flux, with the contraction with the internal tensors appearing in the D7 gaugino coupling understood. Its pointwise scaling is derived in Appendix~\ref{app:g3s}.
Note that the $\tau_s$ prefactor formally following from the integration over the brane stack should not be taken seriously since we do not claim sub-exponential precision in our previous discussion.
Also, we have suppressed the complex conjugate counterpart.

In the Brans-Dicke frame, we then have
\begin{equation}\label{eq:pre_Einstein_linear_potential}
  V^{\rm BD}_{\rm lin} \simeq 
  C_{\rm lin}\,
  \frac{W_0\, \mathrm{e}^{-a\tau_s}}{\sqrt{\mathcal V}}
\end{equation}
and, after Weyl rescaling to the Einstein frame,
\begin{equation}
    V_{\rm lin} \simeq C_{\rm lin}\, \frac{W_0\, \mathrm{e}^{-a\tau_s}}{{\cal V}^{5/2}}\,.
    \label{eq:EFT_linear_volume_scaling}
\end{equation}

The same argument can be repeated for the double-exponential effect: As argued in \cite{Hamada:2021ryq} the D7 brane action contains a quartic gaugino term.\footnote{
This arose in a recent discussion of the 10d derivation of KKLT, including many authors \cite{Moritz:2017xto,Gautason:2018gln, Hamada:2018qef, Kallosh:2019oxv, Hamada:2019ack, Carta:2019rhx, Gautason:2019jwq, Bena:2019mte, Kachru:2019dvo}.
}
The detailed dynamics, including the divergent flux-backreaction effect, the counterterm, and a finite piece, are immaterial for the present paper. At the level of our present parametric analysis, we may simply write
\begin{equation}
    S_{\lambda^4} \;  \sim \;  \int \dif^4x\,\sqrt{-g^{\rm BD}_4}\, c_4(\tau_s)\, {\cal J}_4\; , \qquad {\cal J}_4 \equiv \left\langle (\lambda\lambda)(\bar\lambda\bar\lambda) \right\rangle_{\rm loc}\,,
    \label{eq:quartic_gaugino_source}
\end{equation}
where
\begin{equation}
    {\cal J}_4 \; \sim \;  \Lambda_{\rm strong}^{6}\;  \sim \;  M_s^{6}\mathrm{e}^{-2a\tau_s}\, .
\end{equation}
The induced energy densities before and after Weyl rescaling then read
\begin{equation}
    V^{\rm BD}_{\rm quad} \;\simeq\; C_{\rm quad}\mathrm{e}^{-2a\tau_s}\,\,,\qquad \qquad V_{\rm quad} \;\simeq\; C_{\rm quad} \frac{\mathrm{e}^{-2a\tau_s}}{{\cal V}^{2}}\,.
\label{eq:EFT_quadratic_volume_scaling}
\end{equation}
Combining the linear and quadratic effects,  we have
\begin{equation}
    V_{\mathrm{EFT}} \;  \sim \;  C_{\rm lin} \frac{W_0\, \mathrm{e}^{-a\tau_{s}}}{\mathcal V^{5/2}} + C_{\rm quad}\frac{\mathrm{e}^{-2a\tau_{s}}}{\mathcal V^2}\, .
\end{equation}

Thus, as in the instanton case, the 10d analysis fails to reproduce the volume scaling of the non-perturbative terms in the familiar scalar potential
if one identifies the holomorphic coordinate $\tau_{s,\ho}$ with the volume of the small--cycle.

Finally, we comment on possible threshold corrections to the above result. Since the condensate depends exponentially on the gauge coupling, a logarithmic correction proportional to $\log({\cal V})$ can generate the appropriate modification required to recover the 4d supergravity scaling.

Physically, such corrections admit two complementary descriptions. In the open-string picture, strings ending on the D7--brane probe the neighbourhood of the wrapped divisor. The contribution from those strings that resolve the D7--brane as an extended source in the internal space is expected to be logarithmic,
\begin{equation}
    \Delta_{\rm th}   \; \sim \; \log\left(\frac{\Lambda_{+}(\tau_s)}{M_s}\right) \; \sim \; \log(\tau_s)\, .
    \label{eq:local_D7_threshold}
\end{equation}
Here $\Lambda_{+}(\tau_s)$ denotes the upper scale to which such strings remain sensitive to the local codimension-two geometry.\footnote{This scale is related to the ``winding string scale'' $M_w$ introduced in \cite{Conlon:2009xf,Conlon:2009kt}.} This scale is expected to be determined entirely by the geometry near the wrapped divisor, with no dependence on the bulk volume. In particular, a naive dimensional estimate identifies $\Lambda_{+}$ with the mass of a string stretched over a distance of order the characteristic radius $R_{\tau_s}$ of the divisor. Since the string tension scales as $T_{\rm F}\sim\alpha'^{-1}$, one finds
\begin{equation}
    \Lambda_{+}(\tau_s)  \; \sim \; T_{\rm F}R_{\tau_s}  \; \sim \;\frac{R_{\tau_s}}{\alpha'}  \; \sim \; \tau_s^{1/4}M_s\, .
\end{equation}
A precise determination of this scale lies beyond the scope of our analysis. The only point relevant here is the EFT expectation that it is controlled by local data and carries no dependence on the bulk volume ${\cal V}$.

Equivalently, in the closed-string picture, the same effect can also be described by bulk fields sourced by the D7--brane. Their propagation is effectively two-dimensional only within the local neighbourhood where the wrapped divisor appears flat. This gives rise to a logarithmic profile controlled by $\tau_s$, accounting for the $\log\tau_s$ dependence in \eqref{eq:local_D7_threshold}. At larger distances, the divisor appears localised at a point in the six-dimensional bulk and the profile becomes power-like.

We therefore find no generic EFT mechanism by which this local codimension-two logarithm could extend to the bulk radius and generate a leading $\log{\cal V}$ correction. Such a dependence would require an additional sector whose propagation remains effectively two-dimensional up to the bulk scale.

\subsection{Kaplunovsky--Louis formula and enhanced UV scale}

A complementary formulation of the tension can be given using the  Kap\-lu\-nov\-sky--Louis (KL) formula \cite{Kaplunovsky:1993rd,Kaplunovsky:1994fg} 
relating holomorphic and physical gauge coupling\footnote{For an alternative perspective see \cite{deAlwis:2012bm}. The argument presented there, however, does not affect the conclusions of the EFT analysis in our previous subsection.}
\begin{equation}
\begin{aligned}
\frac{1}{g^2(\mu)} &= \frac{\tau_{s,\ho}}{4\pi} +\frac{\beta_0}{8\pi^2}\log\!\left(\frac{M_P}{\mu}\right)+\frac{c}{16\pi^2}K\\[0.3em]
& \quad- \sum_R\frac{T(R)}{8\pi^2}\log\det Z^R+\frac{T(G)}{8\pi^2}\log 
g^{-2}(\mu)
\,.
\end{aligned}
\label{eq:KLFormula}
\end{equation}
Here $Z^R$ denotes the K\"ahler metric for matter in representation $R$ and 
\begin{equation}
    c = \sum_R n_R T(R) - T(G).
\end{equation}
We attempt to provide some elementary intuition for  \eqref{eq:KLFormula}, avoiding superspace techniques, in App.~\ref{appendix:KLwithoutSuperspace}.

For simplicity, let us assume that no charged matter is present, i.e.,~focus on pure gauge theory. Moreover, being interested in the parametric volume dependence, we can drop the last term $\sim \log g^{-2}$. The KL formula then simplifies to
\begin{equation}
    \frac{4\pi}{g^2(\mu)}= \tau_{s,\ho} +\frac{\beta_0}{2\pi}\log\!\left(\frac{M_P\,\mathrm{e}^{K/6}}{\mu}\right)\,.
\end{equation}
Comparing with
\eqref{eq:D7_log_running}
and assuming $\tau_{s,\ho}\simeq \tau_s$, this implies 
\begin{equation}
    \Lambda^{(\text{KL})}_{\text{UV}} \; \sim \; M_P \mathrm{e}^{K/6} \;  \sim \; \mathcal{V}^{1/6} M_s,
\end{equation}
which clearly clashes with the 10d EFT expectation $\Lambda_{\text{UV}} \sim M_s$.

Threshold corrections associated with an enhanced scale were found in the local orbifold and orientifold models of \cite{Conlon:2009xf,Conlon:2009kt}. In those constructions, specific sectors propagate along a single complex internal direction and therefore remain effectively two-dimensional up to the bulk scale, generating a dependence on $\log{\cal V}$. We do not identify an analogous sector in the present construction and, therefore, see no mechanism by which this effect could arise.

\subsection{Volume-dependent fermion normalisation}\label{sec:Vgaugino}

Before concluding this section, let us consider our findings in view of \cite{Hamada:2021ryq}.\footnote{See also \cite{Moritz:2017xto,Gautason:2018gln, Hamada:2018qef, Kallosh:2019oxv, Hamada:2019ack, Carta:2019rhx, Gautason:2019jwq, Bena:2019mte, Kachru:2019dvo}.
}
A central result of that work is that a completely local 10d action (cf.~their eq.~(40)) reproduces the 4d action underlying KKLT, without appealing to any `supergravity magic' during the derivation.

To be specific, the starting point is the 10d Einstein frame action with the D7 brane fermions canonically normalised in 8d. These are decomposed according to
\begin{equation}
    \Psi=\lambda\otimes\chi\,,
\end{equation}
where $\lambda$ denotes the 4d fermion and $\chi$ the corresponding brane zero mode, normalised by $\chi^\dagger \chi =1$.

In their section~4.3, the external metric is then brought to four-dimensional Einstein frame and the gaugino is correspondingly rescaled to the 4d-SUGRA-normalised field $\lambda_{4d}$ (i.e.~with the inverse gauge coupling multiplying the kinetic term). In our notation, their relation reads
\begin{equation}
    \langle\lambda\lambda\rangle = \sqrt{2}\,c\,{\mathcal K}^{3/2}  \langle\lambda\lambda\rangle_{4d}\, ,
    \label{eq:8D_4d_gaugino_rescaling}
\end{equation}
where ${\mathcal K}=3!\,{\rm Vol}(X)\sim{\mathcal V}$ and $c$ is a numerical constant.\footnote{The same volume scaling can be obtained following the related dimensional-reduction and fermion-normalisation analysis of \cite{Kachru:2019dvo}. There, however, only the single-K\"ahler-modulus case was studied.} The original 10d action is then shown to reproduce the standard four-dimensional supergravity gaugino action, cf.~their eqs.~(53)--(59). 

To extract the non-perturbative scalar potential, one must then additionally use the standard supergravity expectation value
\begin{equation}
    \left\langle\lambda\lambda\right\rangle_{4d} \; \sim \; \mathrm{e}^{K/2}A \mathrm{e}^{-aT_{s,\ho}} \; \sim \; \frac{\mathrm{e}^{-aT_{s,\ho}}}{{\mathcal V}}\, ,
    \label{eq:4d_Sugra_input}
\end{equation}
where $K =-2\log\mathcal{K} + \ldots$ is the four-dimensional K\"ahler potential and the ellipsis stands for terms independent of the K\"ahler moduli. Inserting this expectation value into the 10d-derived 4d action reproduces in their case the KKLT, in our case the (non-perturbative parts of the) LVS scalar potential. 

However, combining \eqref{eq:4d_Sugra_input} with \eqref{eq:8D_4d_gaugino_rescaling} implies
\begin{equation}
    \left\langle\lambda\lambda\right\rangle \; \sim \;  {\mathcal K}^{3/2}\mathrm{e}^{K/2}\mathrm{e}^{-aT_{s,\ho}} \; \sim \; {\mathcal V}^{1/2}\mathrm{e}^{-aT_{s,\ho}}\, .
\end{equation}
Thus, the condensate of fermions with 8d canonical normalisation is enhanced by a factor ${\mathcal V}^{1/2}$.

We see that, on the one hand, the action-level matching of \cite{Hamada:2021ryq} correctly reproduces the 4d supergravity result once the supergravity value of the condensate is supplied. On the other hand, from the perspective of a local D7 brane gauge theory the required volume enhancement of the fermion expectation value of the brane is difficult to understand: as we showed in the previous sections, ordinary gauge running with a local UV scale would instead suggest that $\langle\lambda\lambda\rangle$ remains independent of ${\mathcal V}$ at fixed $\tau_s$. As is particularly clear in this sub-section, the issue is not merely one of the phenomenological LVS model. It concerns, much more broadly, KKLT-type non-perturbative potentials in the regime of parametric scale separation between the volume and smaller cycles.

\section{Log corrections as a possible way out}\label{wout}

As shown in the previous sections, the local EFT description of both E3
instantons and gaugino condensation fails to reproduce the volume dependence of two terms in the LVS potential:
\begin{equation}
V_{\text{EFT}} \; \supset \; \frac{W_0\, \mathrm{e}^{-a \tau_{s}}}{\mathcal{V}^{5/2}} + \frac{\mathrm{e}^{-2a\tau_{s}}}{\mathcal{V}^2} \qquad \quad\text{vs.} \quad\qquad V_{\text{LVS}} \; \supset \; \frac{W_0\, \mathrm{e}^{-a \tau_{s,\ho}}}{\mathcal{V}^{2}} + \frac{\mathrm{e}^{-2a\tau_{s,\ho}}}{\mathcal{V}}\,.
\label{comp}
\end{equation}
Here $\tau_s$ denotes the physical volume of the divisor supporting the non-perturbative effect, while
\begin{equation}
    \tau_{s,\ho}\equiv\mathrm{Re}(T_{s,\ho})
\end{equation}
is the (real part of) the corresponding 4d chiral coordinate. The two expressions are incompatible under the leading-order identification
$\tau_{s,\ho}=\tau_s$. The numerical prefactors and powers of $\tau_{s,\ho}$, suppressed in \eqref{comp}, can not affect this conclusion.

In the present section, we discuss possible corrections that might reconcile 10d EFT and 4d SUGRA, and more generically assess their implications for the robustness of the LVS. Our focus is on logarithmic corrections, inspired by \cite{Conlon:2010ji}.  This should be contrasted with \cite{Klaewer:2020lfg} where, in the class of F-theory models admitting a perturbative heterotic dual, such logarithmic corrections are argued to vanish for certain divisor classes, while they may survive for section-type divisors. Thus the correction invoked here is not automatic: it is compatible with the result of \cite{Klaewer:2020lfg} only if the LVS non-perturbative divisor is of the allowed type, or if the compactification lies outside the assumptions of their heterotic-dual analysis.

\subsection{A two-parameter set of possibilities}
Inspired by \cite{Conlon:2010ji}, we consider a logarithmic correction to the relation between the physical small-cycle volume and the corresponding chiral K\"ahler coordinate
\begin{equation}
    \tau_{s,\ho}    =    \tau_s+\alpha\log(\mathcal V)\,.
    \label{eq:log_redefinition}
\end{equation}
This relation by itself does not determine how the K\"ahler potential entering the supergravity description is modified. 
We therefore allow for an independent correction to the K\"ahler potential that we choose to parametrise as
\begin{equation}
    K+\Delta K = -2\log\left( \mathcal V_{\tilde{\alpha}} +\frac{\xi}{2g_s^{3/2}} \right) + K_{\rm cs} -\log(S+\overline S)\,,
    \label{eq:general-K}
\end{equation}
where
\begin{equation}
    \mathcal V_{\tilde\alpha} = \frac{1}{\lambda} \left[\tau_{b,\ho}^{3/2} -(\tau_{s,\ho}-\tilde\alpha \log(\mathcal{V}))^{3/2}\right]\, .
    \label{eq:general-volume}
\end{equation}
Here $\tilde \alpha$ is independent of the coefficient $\alpha$ entering the field redefinition. 
The physical Calabi--Yau volume is instead
\begin{equation}
    \mathcal V = \frac{1}{\lambda}\left[\tau_{b}^{3/2}-\tau_{s}^{3/2}\right]=\frac{1}{\lambda}\left[\tau_{b,\ho}^{3/2}-\left(\tau_{s,\ho}-\alpha\log(\mathcal V)\right)^{3/2}\right].
    \label{eq:phys-volume-E3-log}
\end{equation}

As a first consistency check, the redefinition can be applied directly to the parametric EFT result. Using \eqref{eq:log_redefinition}, the EFT expression becomes
\begin{equation}
V_{\text{EFT}} \; \sim \; \frac{W_0\, \mathrm{e}^{-a \tau_{s,\ho}}}{\mathcal{V}^{5/2 - a \alpha}} + \frac{\mathrm{e}^{-2a\tau_{s,\ho}}}{\mathcal{V}^{2 - 2a \alpha}} \, .
\label{vmat}
\end{equation}
Agreement with the LVS scaling requires
\begin{equation} \label{eq:correct_alpha}
    \alpha=\frac{1}{2a}\,.
\end{equation}
In more detail, this implies $\alpha=1/4\pi$ for a single $O(1)$ E3 instanton and $\alpha=(N-2)/4\pi$ for pure $\mathrm{SO}(N)$ gaugino condensation. Notice that this matching leaves the parameter $\tilde \alpha$ completely undetermined. 

This parametric matching is only a preliminary check. Since the K\"ahler potential as a function of the holomorphic coordinates is now modified, the complete $F$-term scalar potential must be rederived. Keeping only the leading terms in the large-volume expansion one obtains\,\footnote{Notice that, even for the specific choice $\tilde\alpha=\alpha$ of \cite{Conlon:2010ji}, our expression for the scalar potential differs from Eq.~(4.2) of that paper. The reason is the small-$\alpha$ expansion used therein. A related but distinct scalar potential also appears in \cite{Cicoli:2012fh}. There, the log correction was applied only to an extra blow-up mode appearing quadratically in the Kahler potential. 
}

\begin{equation}
\begin{aligned}
V=&\; \frac{8a^2\lambda |A|^2\sqrt{\tau_{s,\ho}}\, \mathrm{e}^{-2a\tau_{s,\ho}}}{3\mathcal V} \sqrt{1-\frac{\tilde{\alpha}\log(\mathcal V)}{\tau_{s,\ho}}}\\
&\; +\frac{2a\tau_{s,\ho}\,  \mathrm{e}^{-a\tau_{s,\ho}}}{\mathcal V^2} \left(W_0\overline A+A\overline W_0\right) \left[ 1+\frac{3\tilde{\alpha}-2\tilde{\alpha}\log(\mathcal V)}{2\tau_{s,\ho}} \right]\\
&\;+\frac{3\xi |W_0|^2}{4g_s^{3/2}\mathcal V^3}\left[1-\frac{6g_s^{3/2}\tilde{\alpha}\sqrt{\tau_{s,\ho}}}{\lambda\xi} \sqrt{1-\frac{\tilde{\alpha}\log(\mathcal V)}{\tau_{s,\ho}}}\right] \, .
\label{nlvs}
\end{aligned}
\end{equation}

We now study whether the corrected potential still admits an LVS-type minimum at parametrically large volume.
As in standard LVS, one actually has to start with a slightly more general potential, including the axionic partner of $\tau_s$. We assume that the latter has already been integrated out. Absorbing the resulting phase in $W_0\overline{A}$ implies
\begin{equation}
    W_0\overline A+A\overline W_0=-2|AW_0|\,.
\end{equation}
We now assume, to be checked a posteriori, that $\tau_{s, \ho}\sim 1/g_s\gg 1$ (with all other constants ${\cal O}(1)$). As a result, the $\tau_{s, \ho}$-dependent piece of the third term in \eqref{nlvs} becomes sub-dominant. The minimisation w.r.t.~$\tau_{s, \ho}$ involves only the first two terms and, at leading order in $\tau_{s, \ho}$, implies
\begin{equation}
    \mathrm{e}^{-a\tau_{s,\ho}} \simeq \frac{3|W_0|}{4a\lambda|A|} \frac{\sqrt{\tau_{s,\ho}}}{\mathcal V}\sqrt{1-\frac{\tilde\alpha \log(\mathcal V)}{\tau_{s,\ho}}}\,.
\label{eq:modified_taus_eom}
\end{equation}
From this, neglecting sub-leading terms, the standard LVS relation
\begin{equation}
    a\tau_{s,\ho}\simeq\log(\mathcal V)
\label{eq:modified_log_relation}
\end{equation}
follows. Using this result in \eqref{eq:modified_taus_eom}, we obtain the more precise relation
\begin{equation}
    \mathrm{e}^{-a\tau_{s,\ho}} \simeq \frac{3|W_0|}{4\,a\lambda|A|} \frac{\sqrt{\tau_{s,\ho}}}{\mathcal V}\, \sqrt{1-a\tilde\alpha}\, ,
    \label{prer}
\end{equation}
or equivalently
\begin{equation}
    \mathcal V \simeq \frac{3|W_0|}{4\,a\lambda|A|} \sqrt{1-a\tilde\alpha} \, \sqrt{\tau_{s,\ho}}\, \mathrm{e}^{a\tau_{s,\ho}}\,.
\label{eq:modified_volume}
\end{equation}
The existence of this large-volume branch therefore requires
\begin{equation}
    1-a\tilde\alpha >0.
\end{equation}
Substituting \eqref{prer} back into the scalar potential, to leading order, one gets
\begin{equation}
    V_{\mathrm{eff}} \simeq\frac{3|W_0|^2}{4\mathcal V^3 \lambda} \left[ \frac{\lambda \xi}{g_s^{3/2}} -2(1-a\tilde\alpha)^{3/2}\; \tau_{s,\ho}^{3/2} \right]\,.
\label{eq:modified_effective_potential}
\end{equation}
We may now view $\tau_{s,\ho}$ as a function of $\mathcal V$, defined by \eqref{prer}, and minimise in ${\cal V}$. It is easy to understand that, since $\tau_{s,\ho}$ depends on $\mathcal V$ only logarithmically, this amounts to demanding that the square bracket in \eqref{eq:modified_effective_potential} vanishes. Thus, introducing
\begin{equation}
    \tau^{\rm LVS}_s \equiv \frac{1}{g_s} \left(\frac{\lambda\xi}{2}\right)^{2/3}
\end{equation}
as the usual small-cycle value of the LVS, we obtain the log-corrected leading-order result
\begin{equation}
 \tau_{s,\ho} \simeq \frac{1}{1- a\tilde\alpha} \tau^{\rm LVS}_s . 
\label{eq:modified_tau_minimum}
\end{equation}
As in the standard LVS, the leading terms in the scalar potential cancel when evaluated at the minimum, and the AdS vacuum energy is determined by the first subleading contribution:\footnote{This phenomenon, first observed in \cite{Hebecker:2012aw}, was discussed in more detail under the name `non-perturbative no-scale structure' in \cite{Junghans:2022exo}.} $V_{\rm eff}\sim -|W_0|^2\sqrt{\tau_{s, \ho}}/{\cal V}^3$.

Finally, the physical four-cycle volume follows by combining \eqref{eq:modified_tau_minimum}, \eqref{eq:modified_log_relation} and \eqref{eq:log_redefinition}. At leading order,
\begin{equation}
\tau_s\simeq(1-a\alpha)\tau_{s,\ho}\simeq\frac{1-a\alpha}{1-a\tilde\alpha}\tau_s^{\rm LVS}.
\label{eq:modified_positive_cycle}
\end{equation}
Thus the stabilised volume of the small cycle depends on both coefficients, $\alpha$ and $\tilde\alpha$.
Fixing $\alpha=1/(2a)$ as required by the 10d/4d matching, this simplifies to
\begin{equation}
\tau_{s,\ho}\simeq\frac{1}{1-a\tilde\alpha}\tau_s^{\rm LVS},\qquad \tau_s\simeq\frac{1}{2(1-a\tilde\alpha)}\tau_s^{\rm LVS}.
\label{eq:general-matched-minimum}
\end{equation}

\subsection{Concrete proposals and general remarks}

In what follows, we consider two simple prescriptions for the value of $\tilde\alpha$ within this general result. We then briefly discuss possible more general implications.

\paragraph{Preserving the K\"ahler potential as a function of physical volumes:}
As a first option, we follow the prescription proposed in \cite{Conlon:2010ji}. This means that the K\"ahler potential retains its standard form when expressed in terms of physical four-cycle volumes. In our parametrisation, we then have
\begin{equation}
    \tilde\alpha = \alpha,
\end{equation}
so that $\mathcal V = \mathcal V_{\tilde \alpha}$. 
For this choice, the general result \eqref{eq:general-matched-minimum} gives
\begin{equation}
    \tau_{s,\ho} \simeq 2 \tau^{\rm LVS}_s\,, \qquad \tau_s \simeq \tau^{\rm LVS}_s\,.
\end{equation}
Remarkably, the physical four-cycle volume at the minimum coincides with the standard LVS result. Thus, although the logarithmic redefinition produces an order-one shift in the relation between the physical small-cycle volume and the holomorphic variable, the stabilised value of $\tau_s$ is left unchanged at leading order.\footnote{More generally, one may check that for $\alpha = \tilde \alpha$ this property holds for any logarithmic redefinition with $1-a\alpha >0$.}

\paragraph{Preserving the K\"ahler potential as a function of the holomorphic coordinates:}
A second particularly simple possibility is to demand that the K\"ahler potential maintains its form as a function of the holomorphic variable. This corresponds to 
\begin{equation}
    \tilde \alpha = 0\,.
\end{equation}
In this case, the standard LVS scalar potential and its minimisation remain unaffected if one works with the holomorphic variables. Equation \eqref{eq:general-matched-minimum} then gives
\begin{equation}
    \tau_{s,\ho} \simeq  \tau^{\rm LVS}_s, \qquad \tau_s \simeq \frac{1}{2}\tau^{\rm LVS}_s.
\end{equation}
In contrast to the previous case, now the small-cycle physical volume is smaller than in the standard LVS case by a factor of 2.

\paragraph{General remarks:} Our analysis illustrates that, while the logarithmic redefinition can reconcile the parametric 10d and 4d scalings, the stabilised small-cycle physical volume depends on the precise coefficients of the corrections. In particular, the general expression makes it clear that, for sufficiently large $\tilde{\alpha}$,  
the putative minimum can become pathological, for instance because the physical volume is driven to zero or negative values. We see that, once such corrections modify the leading terms in the scalar potential, the fate of the LVS depends on their precise structure and cannot be assessed without an explicit microscopic derivation.

While it is not clear to us which value of $\tilde\alpha$ is preferred, we want to add an observation concerning the choice $\tilde{\alpha}=\alpha$ of \cite{Conlon:2010ji}. One easily checks that, at leading order in the large volume expansion, this choice reproduces the tree-level kinetic term for the physical small-cycle modulus $\tau_s$, consistently with an unperturbed Calabi-Yau geometry. Of course, assuming that the log correction we discuss comes from some explicit $\alpha'$ or loop correction, it is not clear whether this form of the kinetic term should be preserved. However, if one entertains the possibility that no such specific microscopic effect exists and the field redefinition is simply a consistency requirement for achieving a 4d SUGRA description, then this choice appears particularly natural.

Next, as emphasised already in \cite{Conlon:2010ji}, a similar log redefinition of the big cycle $\tau_b$ is structurally much more dangerous for the LVS. However, if one accepts our arguments for the universal presence of the log correction for $\tau_s$, even in the absence of a D7-brane stack, it becomes important to understand the corresponding correction on the full K\"ahler moduli space and, in particular, why it should have no dangerous component along the large-volume direction. 
We do not have an argument that such a correction to $\tau_b$ must be present.
Rather, we want to emphasise that settling the issue of corrections of this type is mandatory if one wants to establish the robustness of the LVS scheme.

Finally, we note that there is a way forward even without quantifying any potential logarithmic field redefinitions. Indeed, one may put the 4d supergravity approach aside and develop the purely 10d analysis. If one can determine the two terms $\sim \exp(-a\tau_s)$ and $\sim \exp(-2a\tau_s)$ including their sub-leading power-like $\tau_s$-dependence, one could directly assess whether their interplay is sufficient to implement LVS moduli stabilisation. We did not achieve this since we derived the non-perturbative terms only at exponential precision in $\tau_s$. It is, however, clear that effects of the required type are induced by the 2-loop $\beta$-function as well as by carefully normalising the fermionic zero modes on the brane. We leave this to future work.

\section{Conclusions}\label{conc}

The purpose of this work was to investigate whether the parametric structure of the scalar potential predicted by 4d supergravity can be recovered directly from a 10d EFT description. While the perturbative $(\alpha')^3$ contribution admits a straightforward 10d interpretation, we found a persistent mismatch for the two non-perturbative terms. Both for E3 instantons and for gaugino condensation, a local ten-dimensional analysis yields
\begin{equation}
    V_{\rm EFT} \; \supset \; \frac{W_0\, \mathrm{e}^{-a\tau_s}}{\mathcal{V}^{5/2}} + \frac{\mathrm{e}^{-2a\tau_s}}{\mathcal{V}^{2}}\, ,
    \label{res10}
\end{equation}
rather than the standard LVS scaling
\begin{equation}
    V_{\rm LVS} \; \supset \; \frac{W_0\, \mathrm{e}^{-a\tau_s}}{\mathcal{V}^{2}} + \frac{\mathrm{e}^{-2a\tau_s}}{\mathcal{V}}\, .
    \label{res4}
\end{equation}
This points to an important limitation in our present understanding of the LVS. Agreement between \eqref{res10} and \eqref{res4} 
requires additional volume dependence on either side. A concrete way to achieve this is through logarithmic corrections to the 4d EFT.
Since, in the LVS regime, these corrections must be comparable to the leading terms, this is a serious concern.

As an illustration, inspired by \cite{Conlon:2010ji} and expanding upon their work, we considered a logarithmic redefinition of the holomorphic variable appearing in the supergravity description 
\begin{equation}
    \tau_{s,\ho} = \tau_s + \alpha \log(\mathcal{V})\, ,
\end{equation}
with $\alpha=1/4\pi$ for a single $O(1)$ E3 instanton and $\alpha=(N-2)/4\pi$ for pure $\mathrm{SO}(N)$ gaugino condensation. 
We allowed in addition for an independent correction to the K\"ahler potential, parametrised by an independent coefficient $\tilde\alpha$. This one-parameter family of models can reconcile the 10d description with the standard LVS scaling and admits an LVS-like AdS minimum provided the correction to the K\"ahler potential is not too large.

On the one hand, it is reassuring that corrections reconciling the 4d and 10d approaches exist
\textit{in principle}. On the other hand, the microscopic mechanism generating them has not been identified and it has not been shown that the coefficients involved take the required value.
This qualification is crucial, as larger corrections could spoil 
the interplay between the different contributions to the 
scalar potential
on which the LVS minimum relies. We emphasise that, while logarithmic corrections have been identified as potentially dangerous for the LVS \cite{Conlon:2010ji, Junghans:2022exo, ValeixoBento:2023nbv}, the option that their coefficient is small or zero has been left open. Our point is very different: We argue that consistency \textit{enforces} the presence of an additional volume dependence of a precise parametric size; if this effect is encoded in a logarithmic redefinition, its coefficient is such that it induces ${\cal O}(1)$ modifications to key relations within the LVS.

The logarithmic effects invoked in this interpretation may originate in $\alpha'$ corrections, loops, or another microscopic source. If these corrections modify the 4d EFT in the desired way, leaving the 10d logic unchanged, consistency can be restored. Establishing this mechanism requires a derivation of both the corrected relation between geometric and chiral variables and its implications for the K\"ahler potential.

Alternatively, the origin of the mismatch may lie on the 10d side, i.e.,~in our understanding of how the local brane dynamics is embedded into the full compactification. Our arguments relied on the assumption that, at small $\tau_s$ and large $\mathcal{V}$, the dynamics of the local QFT living in the small cycle region becomes insensitive at leading order to the global Calabi--Yau geometry. This assumption may break down due to 10d field-theoretic effects we may have missed or, more extremely, because of some form of large-scale stringy non-locality.

We close by emphasising that all of the above should be understood within the general hierarchy relating string theory, 10d EFT and 4d EFT. We do not regard either the 10d or the 4d route as more authoritative. The internal consistency of a given 4d $\mathcal{N}=1$ supergravity model does not by itself establish its realisation as a string compactification. Conversely, the 10d analysis performed in this paper does not represent a fully-fledged string theory calculation and therefore cannot by itself invalidate the standard LVS result. Ultimately, the issue must be fixed in string theory, from which both effective descriptions should emerge and, whenever simultaneously applicable, agree. Until such agreement is achieved, the control of the LVS remains an open question.

More broadly, the issue concerns the emergence of 4d supergravity from non-perturbative dynamics in string compactifications.
It arises independently of the LVS or other phenomenological applications and concerns our quantitative understanding of localised non-perturbative effects in the large volume limit.

\section*{Acknowledgements}We thank Luca Martucci, Fernando Quevedo, Raffaele Savelli, and Liam McAllister for helpful comments and discussion. AS is particularly grateful to Michele Cicoli, Fernando Quevedo, Raffaele Savelli, and Roberto Valandro for earlier collaborations on related topics. 
This work was supported by the Deutsche Forschungsgemeinschaft (DFG, German Research Foundation) under Project Number 552080800. The research of AS is supported by NSF grant PHY-2309456.

\appendix

\section*{Appendix}

\section{The Kaplunovsky–Louis formula without Superspace}\label{appendix:KLwithoutSuperspace}

In this appendix, we show how the Kaplunovsky--Louis (KL) formula \cite{Kaplunovsky:1993rd,Kaplunovsky:1994fg} can be easily understood without invoking the superspace formalism of $\mathcal N=1$ supergravity. The KL formula relates the holomorphic gauge coupling to the physical one in a locally supersymmetric effective field theory:
\begin{equation}
\begin{aligned}
\frac{1}{g^2(\mu)}&=  \mathrm{Re}(f)+\frac{\beta_0}{8\pi^2}\log\!\left(\frac{M_P}{\mu}\right)+\frac{c}{16\pi^2}K\\
& \quad- \sum_R\frac{T(R)}{8\pi^2}\log \det Z^R \ +\frac{T(G)}{8\pi^2}\log g^{-2}(\mu)\, .
\end{aligned}
\end{equation}
where we introduced
\begin{equation}
c=\sum_R n_R\ T(R)-T(G),\qquad \beta_0=\sum_R n_R \ T(R)-3T(G).
\end{equation}
Here $R$ runs over the distinct matter representations, $n_R$ denotes their multiplicity, and $Z^R$ is the corresponding matter metric in flavour space.

To understand the structure of the equation, it is useful to start from the $\mathcal N=1$ supergravity Lagrangian in the superconformal formalism \cite{Kallosh:2000ve,Freedman:2012zz}. The relevant terms for our discussion are
\begin{equation}
\begin{aligned}
\mathcal L_{\text{YM+matter}} \; = & \; \frac{\mathcal N}{6}R+\mathrm{Re}(f_{AB})\left[-\frac14 F^A_{\mu\nu}F^{B\,\mu\nu}-\frac12 \bar\lambda^A \slashed{D}\lambda^B\right]\\
            &\; -\mathcal N_I{}^J\left[(D_\mu\phi^I)(D^\mu\bar\phi_J)+\bar{\chi}_J \slashed{D}\chi^I\right],
\end{aligned}
\end{equation}
where $A,B$ are adjoint indices, $(A_\mu,\lambda)$ are the components of the vector multiplet of the gauge theory, $(\phi,\chi)$ are the scalar and fermionic components of a chiral multiplet transforming in a representation $R$ of the gauge group and $\mathcal N$ is the so--called conformal K\"ahler potential.

Our group-theory conventions are
\begin{equation}
[T^A,T^B]=\I f^{ABC}T^C,\qquad T^AT^A=C_2(R)\,\mathbf 1,\qquad \mathrm{Tr}(T^AT^B)=T(R)\,\delta^{AB}\, .
\label{eq:gauge_generators_conventions}
\end{equation}

For our purposes, it is convenient to choose an explicit gauge fixing to reach the Jordan frame\footnote{We work temporarily in Planck units, $M_P = 1$, and restore $M_P$ below.}
\begin{equation}
\mathcal N=-3\,Y\overline{Y}\,\mathrm{e}^{-K/3}\, ,\qquad Y=\overline{Y}=M_P=1.
\end{equation}
Instead of computing $\mathcal N_I{}^J$ explicitly, it is convenient to expand the Kähler potential around a background with vanishing charged matter fields. Denoting by $n$ the neutral moduli and by $q$ the charged matter fields, one obtains
\begin{equation}
\mathcal N_I{}^J\simeq \mathrm{e}^{-K_0/3}\,Z_I{}^J(n,\bar n).
\end{equation}
The relevant two-derivative terms in the Jordan-frame Lagrangian therefore take the form
\begin{equation}
\begin{aligned}
        \mathcal L_{\text{YM+matter}} \; = & \; -\frac{\mathrm{e}^{-K_0/3}}{2}R+\mathrm{Re}(f_{AB})\left[-\frac14 F^A_{\mu\nu}F^{B\,\mu\nu}-\frac12 \bar\lambda^A \slashed{D}\lambda^B\right]\\
            &\; -\mathrm{e}^{-K_0/3}Z_I{}^J\left[(D_\mu\phi^I)(D^\mu\bar\phi_J)+\bar\chi_J\slashed{D}\chi^I\right].
\end{aligned}
\label{eq:JordanFrame_Lagrangian}
\end{equation}
In what follows, we denote the background K\"ahler potential $K_0$ simply by $K$.

To understand the structure behind the KL formula, consider first a pure gauge theory; to reach Einstein frame one performs the Weyl rescaling
\begin{equation}
g_{\mu\nu}\rightarrow \mathrm{e}^{2\sigma}g_{\mu\nu}\, ,\qquad M_P\rightarrow \mathrm{e}^{-\sigma}M_P, \qquad \sigma=-\frac{K}{6}\, .
\end{equation}
The Weyl transformation is anomalous and therefore shifts the gauge coupling. A simple way to see this is through the running coupling, originally given by
\begin{equation}
\frac{1}{g^2(\mu)}=\mathrm{Re}(f)+\frac{\beta_0}{8\pi^2}\log\!\left(\frac{M_P}{\mu}\right).
\end{equation}
Since the Weyl transformation rescales the UV cutoff as
\begin{equation}
M_P\rightarrow M_P\,\mathrm{e}^{K/6}\, ,
\end{equation}
one then finds
\begin{align}
	\frac{1}{g^2(\mu)}&=\mathrm{Re}(f)+\frac{\beta_0}{8\pi^2}\log\!\left(\frac{M_P\mathrm{e}^{K/6}}{\mu}\right)\nonumber\\[0.3em]
	&=\mathrm{Re}(f)+\frac{\beta_0}{8\pi^2}\log\!\left(\frac{M_P}{\mu}\right)+\frac{\beta_0}{48\pi^2}K\, .
\end{align}
Thus the Weyl anomaly generates the contribution
\begin{equation}
\frac{c}{16\pi^2}K\; ,\qquad  \text{with }\qquad c=-T(G),
\end{equation}
which precisely reproduces the K\"ahler-dependent term in KL.

We now go back to the complete expression, assuming the presence of a single chiral multiplet to simplify the notation. After some trivial algebraic steps employing the relation $3c=\beta_0+2T(R)$, one can recast the KL formula as
\begin{equation}\label{eq:KL_CorrectFactorisation}
	\begin{aligned}
		\frac{1}{g^2(\mu)}\quad = & \quad \mathrm{Re}(f)+\frac{\beta_0}{8\pi^2}\log\!\left(\frac{M_P\mathrm{e}^{K/6}}{\mu}\right)\\
		& \quad -\frac{T(R)}{4\pi^2}\log\!\left(Z^{1/2}\mathrm{e}^{-K/6}\right)+\frac{T(G)}{8\pi^2}\log g^{-2}(\mu)\, .
	\end{aligned}
\end{equation}
Comparing this expression with the supergravity Lagrangian in \eqref{eq:JordanFrame_Lagrangian}, it is clear that the three logarithmic terms are associated with the field redefinitions required to bring the theory to Einstein frame and canonical normalisation: the first arises from the anomalous Weyl rescaling, while the last two are due to the rescaling anomalies associated with the transformations required to canonically normalise the chiral and vector multiplets, respectively.

\section{$G_{(0,3)}$ scaling in the large volume limit}
\label{app:g3s}

In this appendix we show explicitly that \textit{everywhere} on the Calabi--Yau
\begin{equation}
    |G_{(0,3)}| = \frac{|W_0|}{\sqrt{\mathcal V}}\,.\label{gsca}
\end{equation}

The unique, harmonic $(3,0)$-form $\Omega$ of a Calabi--Yau $X$ is covariantly constant: $\nabla_m \Omega_{npq} = 0$. Its norm, defined by
\begin{equation}
    |\Omega|^2 = \frac{1}{3!} \Omega_{mnp}\overline{\Omega}^{mnp}\,,
\end{equation}
is then a constant function on $X$:
\begin{equation}
    0=\nabla_q|\Omega|^2 \quad = \quad \partial_q |\Omega|^2\,.
    \label{eq:Norm_Omega_Const}
\end{equation}
This implies
\begin{equation}
    1= -\I \int_X \Omega \, \wedge \, \overline{\Omega} = \int_X \Omega \, \wedge \, \star_6  \overline{\Omega} = |\Omega|^2 \, \mathcal{V}\,,
\end{equation}
where the first equality is our normalisation convention for $\Omega$ and the second step used the fact that $\star_6 \overline{\Omega} = -\I \,\overline{\Omega} $. Therefore, pointwise on $X$,
\begin{equation}
    |\Omega| \, = \, \frac{1}{\sqrt{\mathcal{V}}}\, .
    \label{eq:Omega_Volume_Scaling}
\end{equation}
Finally, the  GKP superpotential
\begin{equation}
    W_0 \equiv \int_{X}\, G_3 \, \wedge \Omega
\end{equation}
depends only on the component 
$G_{(0,3)}$ of the Hodge-decomposition of $G_3$. This implies
\begin{equation}
    G_{(0,3)} = \I W_0 \overline{\Omega}
\end{equation}
which, combined with \eqref{eq:Omega_Volume_Scaling}, gives the desired result \eqref{gsca}.

\bibliographystyle{JHEP.bst}
\bibliography{bibliography}
\end{document}